\documentclass[DIV=calc, paper=a4, fontsize=11pt, twocolumn]{scrartcl}
\pdfoutput=1

\def\signed #1{{\leavevmode\unskip\nobreak\hfil\penalty50\hskip2em
  \hbox{}\nobreak\hfil(#1)%
  \parfillskip=0pt \finalhyphendemerits=0 \endgraf}}

\newsavebox\mybox
\newenvironment{aquote}[1]
  {\savebox\mybox{#1}\begin{quote}}
  {\signed{\usebox\mybox}\end{quote}}

\usepackage{hyphenat}
\usepackage{url}
\usepackage{verbatim}

\usepackage[english]{babel}
\usepackage[protrusion=true,expansion=true]{microtype}
\usepackage{amsmath,amsfonts,amsthm}
\usepackage[svgnames]{xcolor}
\usepackage[hang, small,labelfont=bf,up,textfont=it,up]{caption}
\usepackage{booktabs}
\usepackage{fix-cm}
\usepackage{sectsty}
\allsectionsfont{\usefont{OT1}{phv}{b}{n}}

\usepackage{fancyhdr}
\usepackage{lastpage}

\renewcommand{\headrulewidth}{0.0pt}
\renewcommand{\footrulewidth}{0.4pt}

\usepackage{lettrine}
\newcommand{\initial}[1]{
\lettrine[lines=3,lhang=0.3,nindent=0em]{
\color{DarkGoldenrod}
{\textsf{#1}}}{}}

\usepackage{titling}

\newcommand{\HorRule}{\color{DarkGoldenrod} \rule{\linewidth}{1pt}}

\usepackage{lmodern}

\pretitle{\vspace{-30pt} \begin{flushleft} \HorRule \fontsize{50}{50} \usefont{OT1}{phv}{b}{n} \color{DarkRed} \selectfont}

\title{MISRA C, \\ for Security's Sake!\textsuperscript{$\dagger$}}

\posttitle{\par\end{flushleft}\vskip 0.5em}

\preauthor{\begin{flushleft}\large \lineskip 0.5em \usefont{OT1}{phv}{b}{sl} \color{DarkRed}}

\author{Roberto Bagnara, }

\postauthor{\footnotesize \usefont{OT1}{phv}{m}{sl} \color{Black}
BUGSENG\footnotemark[1]\ \ and University of Parma\footnotemark[4]\ \,
member\footnotemark[3]\ \ of MISRA C Working Group, \\
member\footnotemark[3]\ \ of ISO/IEC JTC1/SC22/WG14 - C Standardization Working Group \\
\vspace{2mm}
\vspace{-1mm}

\par\end{flushleft}\HorRule}

\date{}

\begin{document}
\renewcommand*{\thefootnote}{\fnsymbol{footnote}}

\maketitle

\thispagestyle{fancy}

\initial{A}\textsf{\textbf{\nohyphens{%
third of United States new cellular subscriptions in Q1 2016 were for cars.
There are now more than 112 million vehicles connected
around the world.  The percentage of new cars shipped with Internet
connectivity is expected to rise from 13\% in 2015 to 75\% in 2020, and
98\% of all vehicles are likely to be connected by 2025.  Moreover, the
news is often reporting about ``white hat'' hackers intruding on car
software.  For these reasons, security concerns in automotive and
other industries have skyrocketed.
We briefly illustrate the relationship between
the ISO/IEC C language standards,
the CERT C coding standards,
ISO/IEC TS 17961 (``C Secure Coding Rules''), and MISRA C,
with a focus on the objective of preventing security vulnerabilities
(and of course safety hazards), as opposed to trying to eradicate them
once they have been inserted in the code.
We then introduce two new MISRA documents, MISRA C:2012 Addendum~2 and
MISRA C:2012 Amendment 1, which clarify and complete the coverage offered
by MISRA C:2012 against TS 17961.
These new developments ensure that MISRA C, which is widely respected
as a safety-related coding standard, is equally applicable as a
security-related coding standard.}}}

\footnotetext[2]{Presented at the \emph{14\textsuperscript{th} Workshop on
  Automotive Software \& Systems}, Milan, November 10, 2016.}
\footnotetext[1]{\url{roberto.bagnara@bugseng.com}}
\footnotetext[4]{\url{bagnara@cs.unipr.it}}
\footnotetext[3]{Writing and speaking in a personal capacity.}
\renewcommand*{\thefootnote}{\arabic{footnote}}
\setcounter{footnote}{0}
\newpage

\section*{Joy and Pain of Connected Cars}

Once upon a time, \emph{embedded} systems were \emph{isolated} systems.
This is no longer the case, at least in the automotive industry.
In the first quarter of 2016, connected cars accounted for around a third
of all new cellular subscriptions, more than phones, more than tablets.%
\footnote{\url{http://chetansharma.com/usmarketupdateq12016.htm}}
The percentage of new cars shipped with Internet connectivity is expected
to increase rapidly, from $13\%$ in 2015 to $75\%$ in 2020 to $~100\%$
in 2025, when $98\%$ of all vehicles will likely be connected.
What happens to connected computers is well known and,
according to the
\emph{Common Vulnerabilities and Exposures} (CVE) database,
the situation is not improving despite the much increased awareness
of security issues.  For instance, \emph{denial of service} vulnerabilities
listed in the CVE have been increasing from 1999 up to and including 2016.
The fear that car owners worldwide might become the target of various
kinds of criminals is concrete.  With the number of victims of
ransomware running to hundreds of thousands and the consequent global
losses likely to reach hundreds of millions of euros,
it is conceivable that locking up cars until the owners pay
might be seen as a profitable thing to do.

\begin{aquote}{Tony Lee, FireEye}
  ``For well-organized attackers, this may end up being a numbers game,
    which may be similar to credit-card theft and sale.''
\end{aquote}

\section*{Safety and Security}

The fact that the English words \emph{safety} and \emph{security} correspond to
the same word in many languages
is not helping to clarify the
distinction between the two concepts.
Commonly used taxonomies
define
\(
  \textit{Safety}
    = \textit{Integrity} + \textit{Absence of catastrophic consequences}
\)
and
\(
  \textit{Security}
    =  \textit{Confidentiality} + \textit{Integrity} + \textit{Availability}
\),
where $\textit{Integrity}$ can be described as the absence of improper
(i.e., out-of-spec) system alterations under normal and exceptional conditions
\cite{AvizienisLRL04}.
The only thing that distinguishes the role of integrity in safety and security
is the notion of \emph{exceptional condition}.
This reflects the fact that exceptional conditions are perceived
as accidental (safety hazards) or intentional (security threats)
\cite{GoertzelF09}.

While safety and security are distinct concepts, when it comes to connected
software not having one implies not having the other.
To start with, both safety and security issues are due to software defects:
some are introduced in requirements, some in design, some in coding.
It is well known that many software defects that have an impact
on safety, e.g., buffer overflows, can be exploited to attack a system
if exposed to the outside world: we can summarize that by
\(
  \textit{Unsafe} + \textit{Connected} \implies \textit{Insecure}.
\)
On the other hand, the recent successful attempts of
Charlie Miller and Chris Valasek at impacting safety functions
of the Jeep Cherokee show that
\(
  \textit{Insecure} \implies \textit{Unsafe}
\).
\begin{aquote}{John Carlin, U.S.\ Dept.\ of Justice}
  ``They want to drive trucks into civilians, and it's not too much
    to think they can hack a car and do the same thing.''
\end{aquote}

\section*{C, CERT C, TS~17961, and MISRA C}

ISO/IEC JTC1/SC22/WG14, a.k.a. the \emph{C Standardization Working Group},
has always been faithful to the original spirit of the language
\cite{Seacord16}.
With some little humor, this can be captured in the form of ``commandments''
like the following:
\newcounter{cmdmnt}
\begin{list}%
  {\Roman{cmdmnt}}{\usecounter{cmdmnt}\setlength{\rightmargin}{\leftmargin}}
\item
  Trust the programmer
\item
  Let the programmer do anything
\item
  Keep it fast, even if not portable
\item
  Keep it small and simple
\end{list}
Of course, some of these conflict with both safety and security
requirements.
All that is well known, as is well known that, for safety-related
applications, language subsetting is crucial.
The most authoritative language subset for the C programming language
is MISRA~C, now at its third edition \cite{MISRA-C-2012},
MISRA~C:2012 or MC3 for short.

One of the unfounded myths about MISRA~C is that it is only about
safety.  In reality, in its opening paragraph MISRA~C presents itself
as ``a subset of the C language [that] can also be used to develop
any application with high integrity or high reliability requirements.''
Because of this misunderstanding, when awareness of security threats
increased, people started looking elsewhere for a security-related
C coding standard, and they found \emph{The CERT C (Secure) Coding Standard},
whose first edition was published in 2008 \cite{CERT-C-2008}
and second edition in 2014 \cite{CERT-C-2014}.
Both editions were authored by Robert C.~Seacord when he was employed
by the CERT Division of the Software Engineering Institute at CMU;
Addison-Wesley owns the copyright~\cite{Seacord16}.

Despite its popularity, CERT~C has several shortcomings from an industrial
point of view: it is the product of essentially one person (although
a very expert and talented one, who, however, has now left the project);
its development continues on a shared wiki that changes overnight;%
\footnote{A new ``snapshot version'' has been published in PDF form
  on June 30, 2016 \cite{CERT-C-2016}.}
it is not based on the idea of language subsetting, hence, differently
from MISRA~C, it leans on the \emph{cure} side  rather than
the \emph{prevention side};
many rules are formulated in a way that is not directly amenable
to automatic, static analysis.

While working at what has become the C11 edition of the ISO/IEC C language
standard, WG14 ---which, starting from 2006, encouraged the development
of CERT~C--- established, in 2009, a study group whose objective was
to produce statically analyzable secure coding guidelines for the C language.
The result of this effort has been the publication of the
ISO/IEC TS 17961:2013 technical specification \cite{ISO-IEC_DTS_17961-2013},
TS~17961 in the sequel.

\begin{aquote}{TS 17961, Introduction}
``An essential element of secure coding in the C programming language is
a set of well-documented and enforceable coding rules. The rules
specified in this Technical Specification apply to analyzers,
including static analysis tools and C language compiler vendors that
wish to diagnose insecure code beyond the requirements of the language
standard. All rules are meant to be enforceable by static analysis.''
\end{aquote}

\section*{MC3 for Safety \emph{and} Security}

As we saw, until and including the first quarter of 2016,
the situation was the following:
on the one hand, we had MISRA C:2012, which was and is widely respected
as a safety-related coding standard (even though its prescriptions
go beyond safety and are targeted at all high integrity and high
reliability systems);
on the other hand, we had TS~17961, a security-related coding
standard backed by ISO.
Even though, as mentioned, the software defects that can give
rise to safety hazards and security threats have a significant
intersection, MISRA C:2012 and TS~17961, as they were published in 2013,
are not a substitute for one another.

Now, since the first quarter of 2016, things have changed.
The MISRA~C Working group has published two
documents: the first is \emph{MISRA C:2012 Addendum 2}
\cite{MISRA-C-2012-Addendum2}, which contains
a coverage matrix of MISRA~C:2012 against TS~17961.
The second document, \emph{MISRA C:2012 Amendment~1},
contains $14$ additional guidelines,
$1$ directive and $13$ rules, targeted at the prevention
of security issues~\cite{MISRA-C-2012-Amendment1}.\footnote{Both
documents are available at \url{http://misra.org.uk/} .}
These concern, with `D' standing for \emph{directive}
and `M'/`R' standing for \emph{mandatory/required rule}, respectively:
\begin{description}
\item[1 D]
  validation of external data;
\item[1 M]
  use of \verb+sizeof()+ on a function parameter of array type;
\item[1 M]
  \verb+<ctype.h>+ functions;
\item[3 R]
  \verb+<stdlib.h>+ memory comparison functions;
\item[2 M]
  \verb+<stdlib.h>+ environment functions;
\item[2 M]
  \verb+<string.h>+ string-handling functions;
\item[1 R]
  \verb+<stdio.h>+ I/O functions and handling of \verb+EOF+;
\item[3 R]
  \verb+<error.h>+ handling of \verb+errno+.
\end{description}
The coverage of MISRA~C:2012, without (MC3) and with Amendment~1
(MC3 + MC3A1) is shown in Table~\ref{tab:TS-17961-coverage}.
The coverage kind column has to be interpreted as follows
\cite{MISRA-C-2012-Addendum2}:
\begin{description}
\item[Explicit]
  The behaviour addressed by the TS~17961 rule is \emph{explicitly}
  covered by one or more MISRA C:2012 guidelines, which directly addresses
  the undesired behaviour.
\item[Implicit]
  The behaviour addressed by the TS~17961 rule is \emph{implicitly}
  covered by one or more MISRA C:2012 guidelines, although the behaviour
  is not explicitly referenced.
\item[Restrictive]
  The behaviour addressed by the TS~17961 rule is covered by one
  or more MISRA C:2012 guidelines that prohibit a language feature
  in a \emph{restrictive} manner.
  For example:
  \begin{description}
    \item[Rule 21.3] \verb+stdlib.h+: memory alloc./dealloc.;
    \item [Rule 21.5] \verb+signal.h+: all;
    \item[Rule 21.6] \verb+stdio.h+: input/output functions;
    \item[Rule 21.8] \verb+stdlib.h+: \verb+getenv()+.
  \end{description}
\item[Broad]
  Some aspects of the behaviour addressed by the TS~17961 rule are covered
  in a \emph{restrictive} manner; some other aspects of the behaviour
  are not covered by any MISRA C:2012 guidelines.
\item[None]
  The behaviour addressed by the TS~17961 rule is not covered by any MISRA C
  guidelines.
\end{description}

\begin{table}
\caption{MISRA~C:2012 coverage of TS 17961}
\label{tab:TS-17961-coverage}
\centering
\begin{tabular}{lrr}
\toprule
Coverage kind & MC3 & MC3 + MC3A1 \\
\midrule
Full, explicit    & 22 & 35 \\
Full, implicit    &  7 &  3 \\
Full, restrictive & 11 &  8 \\
Partial, broad    &  2 &  0 \\
None              &  4 &  0 \\
\midrule
Total             & 46 & 46 \\
\bottomrule
\end{tabular}
\end{table}
Table~\ref{tab:TS-17961-coverage} shows that, while coverage for
freestanding applications was already very good before the issue of
MISRA C:2012 Amendment~1, now coverage can be considered complete.

It is also interesting to see how MISRA C:2012 integrated with Amendment~1
covers CERT~C.  The MISRA~C working group is currently working on
the production of a document containing a proper compliance matrix.
Preliminary, unofficial data can be seen in
Table~\ref{tab:CERT-C-coverage}, both for the latest book
edition, CERT C:2014 \cite{CERT-C-2014}, and the PDF snapshot
released on June 30, 2016 \cite{CERT-C-2016} (in which two rules have
been added and one has been deleted).
\begin{table}
\caption{MC3+MC3A1 coverage of CERT C}
\label{tab:CERT-C-coverage}
\centering
\begin{tabular}{lrr}
\toprule
Coverage kind & CERT C:2014 & CERT C:2016 \\
\midrule
C11 specific   & 13 & 14 \\
Full, explicit & 41 & 42 \\
Full, implicit & 17 & 17 \\
Full, restrictive & 22 & 21 \\
None           &  5 &  5 \\
\midrule
Total          & 98 & 99 \\
\bottomrule
\end{tabular}
\end{table}

There are other reasons why MISRA~C:2012 with Amendment~1 is the best
available coding standard for the development of critical embedded
systems.  One of them is the emphasis its guidelines put on readability:
it is well known that code review combined with static analysis and
the automatic enforcement of sound coding guidelines by means
of high-quality tools is, by far, the most effective defect removal strategy.
This is even more so when security concerns regarding, e.g.,
confidentiality or privilege escalation: while progress is being made
on formal methods that are able to address them, careful code review
is crucial both for today and for the foreseeable future.

\begin{aquote}{TS 17961, Introduction}
  ``In practice, then, security-critical and safety-critical code
  have the same requirements.''
\end{aquote}

\section*{Conclusion}

Connected cars are with us and in less than a decade they will be
everywhere.  Given the amount of software that equips them and the
criticality of the functions controlled by it, connectivity opens
the door to malicious activities of all kinds.
This comes at a time when a redefinition of the concept of liability
for producers of embedded software is taking place (cf.\ the Toyota
unintended acceleration case).
When it comes to security, additional issues arise.
For instance, in order to demand a ransom, it is not even necessary
to lock up or compromise safety of the vehicle:
it is enough to, e.g., display on the dashboard something that makes
the owner suspect safety might have been impacted by hackers.
As another example, the security risk posed by disgruntled/unfaithful
developers will likely affect the way software is produced and
verified.
The automotive industry, which already was the most critical
sector for software safety, is also becoming to play the same
role in regard to software security.

In this short essay, we recalled what makes safety and security
different and what is common to them.
We then reviewed the genesis of the CERT~C coding standards and
ISO/IEC TS~17961 ``C Secure Coding Rules'',
and their relationship with the ISO/IEC C language standard.
We discussed recent developments of the MISRA C:2012 coding guidelines,
which complete the coverage of TS~17961.
We argue that, with this integration, MISRA C:2012 is the C coding standard
of choice for the automotive industry and for all industries developing
embedded systems that are safety-critical and/or security-critical.

\begin{aquote}{Kevin Tighe, Bugcrowd}
  ``Car companies are finally realising that what
    they sell is just a big computer you sit in.''
\end{aquote}

\end{document}